\documentstyle[12pt]{article}

\newcommand{\Od}{{\cal O}}

\newcommand{\Dbar}{\not{\!{\!D}}}

\input epsf

\newcommand{\AmS}{{\protect\the\textfont2
  A\kern-.1667em\lower.5ex\hbox{M}\kern-.125emS}}

\begin{document}


\title{Gravitino production during preheating\footnote{Talk given in the International
Workshop on Particles in Astrophysics and Cosmology, Valencia (Spain), May 3-8, 1999.
To appear in the proceedings.} } 
\author{Antonio L. Maroto\thanks{Present address: Dept. F\'{\i}sica Te\'orica, 
Universidad Complutense de Madrid, 28040, Madrid, Spain. E-mail: maroto@eucmax.sim.ucm.es}\\
Astronomy Centre, University of Sussex,  \\
               Falmer, 
               Brighton,
               BN1 9QJ, U.K. }

\date{\today}
\maketitle
\begin{abstract}
We study the production of gravitinos during the preheating era after inflation by
means of the non-perturbative Bogolyubov technique. Considering only the helicity
$\pm 3/2$ states, the problem is reduced to the simpler Dirac fermions case. We calculate 
the production in a particular supergravity model in an expanding universe and 
obtain the spectrum and number density. Finally we compare the results
with the nucleosynthesis bounds and extract some consequences.
\end{abstract}

\section{Introduction}

The presence of weakly interacting massive particles can
 have important consequences
in cosmology \cite{Kolb}. In particular,  when such particles are light, with life times 
longer than the universe
age, they can act as  dark matter or even dominate the present energy density
and overclose the universe. On the other hand, if they are unstable and
decay during the nuclesynthesis era, they could destroy the nuclei 
created in this period and spoil the successful
predictions of the standard big bang model. 

One of such particles is the 
gravitino, i.e, the spin $3/2$ superpartner of the graviton in supergravity theories, 
whose couplings to the rest of particles are suppressed by the Planck mass scale.
Typically its decay rate is given by $\Gamma_{3/2}\simeq m_{3/2}^3/M_P^2$
which implies that gravitinos lighter than $m_{3/2}<100 MeV$ will live longer
than the universe age. 
 Due to their weak couplings, gravitinos
freeze-out very early when they are still relativistic and 
therefore their primordial abundance can be estimated as $n_{3/2}/s\simeq 10^{-3}$
\cite{Kolb}.
This allows us to obtain the well-known bound on the mass of stable gravitinos, 
i.e, in order not to overclose the universe we need 
$m_{3/2}<1keV$ \cite{Pagels}. However if they are unstable, that primordial abundance
would give rise to an enormous amount of entropy that will conflict with the standard
cosmology. A possible way out of 
this {\it gravitino problem} is the existence of a period of inflation that
dilutes any primordial density \cite{LiEll}.
Unfortunately the problem can be recreated if after inflation, gravitinos are produced
by some mechanism. In fact, this could be the case if during the 
period of inflaton oscillations at the end of inflation, the reheating  temperature
was sufficiently high.
A successful nucleosynthesis period requires (we give some conservative
bounds \cite{Sarkarrep}): 
$n_{3/2}/s<10^{-15}$ for a gravitino  mass $m_{3/2}\simeq 100GeV$,
$n_{3/2}/s<10^{-14}$ for $m_{3/2}\simeq 1TeV$ and $n_{3/2}/s<10^{-13}$
for $m_{3/2}\simeq 10TeV$. 
The production of gravitinos during reheating depends on the
temperature $T_R$ as \cite{Ellis}:
 $n_{3/2}/s\simeq 10^{-14}T_R/(10^9 GeV)$, 
which implies that for a typical  mass $m_{3/2}\simeq 1 TeV$, the
reheating temperature has to be $T_R<10^9 GeV$.
Another constraint appears in supergravity models, where
the gravitino mass is determined by the scale of 
supersymmetry breaking. In order to solve the hierarchy problem, it is then suggested 
that $m_{3/2}<1 TeV$ \cite{Bailin}.

As we have just mentioned, during the reheating period gravitinos can be created
by the perturbative decay of other particles produced from the inflaton oscillations.
However, in recent years, the standard picture of reheating has changed
dramatically as a consequence of some works in which it was realized that 
during the first inflaton oscillations, reheating
can not be studied by the standard perturbative
techniques \cite{Linde,brandenberger}.
This so called {\it preheating} period can give rise to an explosive production
of bosons due to the phenomenon of parametric resonance. In this period, the energy
of the coherent oscillations of the  inflaton field is very efficiently 
converted into particles. In the
case of fermions, the limit imposed by Pauli exclusion principle avoids the
explosive production, 
but however the results deviate from the perturbative
expectations \cite{Baacke,GK,MaMa}. 
This fact will be of the utmost importance in the case that gravitinos
are directly coupled to the inflaton,
since the new preheating period can give rise to an excess of production. 

In this work we present our results on the production of spin $3/2$ particles
from the non-perturbative  Bogolyubov technique \cite{paper} and apply the method to
the preheating period in a particular supergravity model.

\section{Quantization of spin $3/2$ fields in external backgrounds}
 
Preheating is based on the phenomenon of creation of particles from classical
backgrounds (see \cite{Mostepanenko} and references therein). 
The spin $3/2$ case is however special due to 
the consistency problems that avoid the  quantization of 
such fields
in the presence of scalar, electromagnetic or gravitational backgrounds \cite{Velo}. 
The only theory in which these problems seem to be absent is supergravity,
provided the background fields satisfy the corresponding equations of motion
\cite{SUGRA}.
But, the complicated form of the Rarita-Schwinger (RS) equation
makes it very difficult to extract explicit results even in
simple cases.  
Let us start by considering the massive RS equation
in a curved space-time. We will include the coupling to a scalar field 
$\chi$ by modifying 
the mass term, (we will follow the notation in \cite{Moroi}):
\begin{eqnarray}
\epsilon^{\mu\nu\rho\sigma}\gamma_5\gamma_\nu D_\rho\psi_\sigma+
\frac{1}{2}(m_{3/2}-\chi)[\gamma^\mu,\gamma^\nu]\psi_\nu=0.
\end{eqnarray}
As usual in supergravity models we will consider Majorana spinors
satisfying $\psi_\mu=C\bar\psi_\mu^T$ with $C=i\gamma^2\gamma^0$ the
charge conjugation matrix.
We have introduced the curvature of space-time
by {\it minimal} coupling as done in supergravity, i.e, 
$D_\rho \psi_\sigma=(\partial_\rho+\frac{i}{2}
\Omega^{ab}_\rho\Sigma_{ab})\psi_\sigma$ with $\Omega^{ab}_\rho$ the 
spin-connection coefficients and $\Sigma_{ab}=\frac{i}{4}[\gamma_a,\gamma_b]$.
The $\epsilon^{\mu\nu\rho\sigma}$ removes the Christoffel symbols
contribution in the covariant derivative.
Following the analogy with the creation of gravitons from  Einstein
equations,
we will consider only the linearized equation in $1/M$ 
($M_P^2=8\pi M^2)$ for supergravity  
\cite{Bailin}, i.e, we will consider only the symmetric part of the 
spin-connection, ignoring the torsion contribution that is $\Od(M^{-2})$.
In flat space-time and when 
$\chi=0$, the general solution of these equations can be expanded in four helicity 
$l=s/2+m$ modes ($l=\pm 3/2, \pm 1/2$):
\begin{eqnarray}
\psi^{pl}_\mu(x)=e^{-ipx}\sum_{s,m} J_{sm}u(\vec p,s)\epsilon_\mu(\vec p,m)
\label{planewaves}
\end{eqnarray}
with $J_{sm}$ the Clebsch-Gordan coefficients whose values are: 
$J_{-1-1}=J_{11}=1$, $J_{-11}=J_{1-1}=1/\sqrt{3}$ and 
$J_{-10}=J_{10}=\sqrt{2/3}$. $u(\vec p,s)$ are spinors with definite 
helicity $s=\pm 1$ and $\epsilon_\mu(\vec p,m)$ with $m=\pm 1,0$ are
the three spin $1$ polarization vectors.

In order to study the preheating era, we need to consider the presence
of the inflaton field and the expanding universe.
With that purpose we take a scalar field only depending on time $\chi(t)$
and a spatially flat Friedmann-Robertson-Walker metric.

Now, the expression in (\ref{planewaves}) 
is not a
solution of the equations of motion. However, we can look 
for general homogeneous solutions of the RS equation
in the form:
\begin{eqnarray}
\psi^{pl}_{\mu}(x)=e^{i\vec p \cdot \vec x}f^{pl}(t)\sum_{s,m}J_{sm}
u(\vec p, s)
\epsilon_\mu(\vec p,m)
\label{ansatz}
\end{eqnarray}
These fields satisfy the condition $\gamma^\mu\psi_\mu=0$
and if we constraint
ourselves to the helicity $l=\pm 3/2$ states, then 
the 
RS equation reduces to a Dirac form \cite{paper}:
\begin{eqnarray}
(i\Dbar -m_{3/2}+\chi)\psi_\mu=0 \label{cons0}
\end{eqnarray}
and the condition  
$D^\mu \psi_\mu=0$  also holds.
As far as these modes satisfy a Dirac-like equation, it suggests that 
all the difficulties in
the gravitino quantization procedure  would reduce to the 
helicity $\pm 1/2$ modes in this case. 
In fact the 
above ansatz (\ref{ansatz})
is not a solution for the helicity $\pm 1/2$ modes even for homogeneous 
backgrounds.

Thus we see that the production of helicity $\pm 3/2$ gravitinos during preheating
in an expanding universe can be studied in the same way as the 
production of Dirac fermions.
With this purpose we have to reduce equation 
(\ref{cons0}) to a second order equation. Let us first write the equation in
conformal time defined as $dt=a(\eta)d\eta$:
\begin{eqnarray}
\left(ia^{-1}\gamma^\mu\partial_\mu-m_{3/2}+\chi
+i\frac{3}{2}\frac{\dot a}{a^2}\gamma^0\right)\psi_\mu=0
\label{dirconf}
\end{eqnarray}
where $\dot a=da/d\eta$. We will take the following ansatz on the
helicity  $l=\pm 3/2$ solutions:
\begin{eqnarray}
\psi_\mu^{pl}(x)=a^{-3/2}(\eta)e^{i\vec p \cdot \vec x} U_\mu^{\vec p l}(\eta)
\end{eqnarray}
with 
\begin{eqnarray}
U_\mu^{\vec p l}(\eta)=\frac{1}{\sqrt{\omega+m_{3/2}^0}}\left(
i\gamma^0\partial_0
-\vec p\cdot \vec\gamma\right.\nonumber \\
 +\left. a(\eta)(m_{3/2}-\chi(\eta)\right)f_{pl}(\eta)u(\vec p, s)
\epsilon_\mu(\vec p, m)
\end{eqnarray}
and the normalization 
$U_\mu^{\vec p l\dagger} (0)U^\mu_{\vec p l}(0)=2\omega$ where  
$m^0_{3/2}=a(0)m_{3/2}$.
It is possible to check that this ansatz automatically satisfies 
$\gamma^\mu \psi_\mu=0$ and $D^\mu \psi_\mu=0$. An appropriate form for
the spinor $u(\vec p, s)$ and polarization
vectors $\epsilon_\mu(\vec p, m)$ can be obtained if we choose 
the Dirac representation
for the gamma matrices and we take (without loss of generality) 
the $z$-axis along the $\vec p$ 
direction. In this case $u(\vec p,1)^T=(1,0,0,0)$, $u(\vec p, -1)^T=(0,1,0,0)$,
$\epsilon_\mu(\vec p, 1)=\frac{1}{\sqrt{2}}(0,1,i,0)$ and
$\epsilon_\mu(\vec p, -1)=\frac{1}{\sqrt{2}}(0,1,-i,0)$. With this choice
$u(\vec p,\pm 1)$ are eigenstates of $\gamma^0$ with eigenvalues $+1$.
Then equation (\ref{dirconf}) reduces to the well-known form:
\begin{eqnarray}
\left(\frac{d^2}{d\eta^2}\right.&+&\left.p^2-i\frac{d}{d\eta}\left(
a(\eta)(m_{3/2}-\chi(\eta)\right)\right. \nonumber \\
&+&\left.
a^2(\eta)(m_{3/2}-\chi(\eta))^2\right)f_{pl}(\eta)=0
\label{master}
\end{eqnarray}
In order to quantize the modes we will expand an arbitrary solution
with helicity $l=\pm 3/2$ as:
\begin{eqnarray}
\psi^{l}_\mu(x)=\int \frac{d^3p}{(2\pi)^3 2\omega}a^{-3/2}(\eta)
\left(e^{i\vec p\cdot \vec x}U_\mu^{\vec p l}(\eta)a_{\vec p l}
+
e^{-i\vec p\cdot \vec x}U_\mu^{\vec p l C}(\eta)a_{\vec p l}^\dagger\right)
\end{eqnarray}
where the creation and annhilation operators satisfy the anticommutation
relations $\{a_{\vec p l},a_{\vec p' l'}^\dagger\}
=(2\pi)^3 2\omega \delta_{ll'}\delta(\vec p-\vec p')$.

\section{A supergravity inflation example}

We will consider a specific
supergravity inflationary model (see \cite{Sarkar}), in which the inflaton
field is taken as the scalar component of a chiral superfield, and its
potential is derived from the superpotential $I=(\Delta^2/M)(\phi-M)^2$. 
This is the simplest choice that satisfies 
the conditions that supersymmetry
remains unbroken in the minimum of the potential and that the present
cosmological constant is zero. CMB anisotropy fixes the inflationary scale
around $\lambda\equiv \Delta/M\simeq 10^{-4}$. For the sake of
simplicity, we will consider the case in which the gravitino mass is
much smaller than the effective mass of the inflaton in this model, 
$m_{3/2}\ll m_{\phi}\simeq 10^{-8}M$ and since the production
will take place during a few inflaton oscillations, 
we will neglect the mass term in 
the equations.
The scalar field potential is shown to be stable in 
the imaginary direction  and
therefore we will take for simplicity a real inflaton field. Along the
real direction the potential can be written as:
\begin{eqnarray}
V(\phi)&=&\lambda^4 e^{\phi^2}\left((2(\phi -1)+\phi(\phi-1)^2)^2-3(\phi-1)^4
\right)
\end{eqnarray}
where we are working in units $M=1$. This potential has
a minimum in $\phi=1$. The coupling of the inflaton field
to gravitinos is given by the following term in the supergravity
lagrangian \cite{Bailin}:
\begin{eqnarray}
{\cal L}_{int}=-\frac{1}{4}e^{G/2}\bar \psi_\mu
[\gamma^\mu,\gamma^\nu]\psi_\nu\\
e^{G/2}=\lambda^2 e^{\phi^2/2}(\phi-1)^2
\end{eqnarray}
where the K\"ahler potential has been chosen in such a way that the 
kinetic terms for the scalar fields are canonical $G(\Phi,\Phi^\dagger)=
\Phi^\dagger \Phi+\log \vert I \vert ^2$.
The inflaton and Friedmann equations can be
written in conformal time as:
\begin{eqnarray}
\ddot \phi +2 \frac{\dot b}{b}\dot \phi +\frac{b^2}{\lambda^4}V_{,\phi}=0\\
\frac{\dot b^2}{b^2}=\frac{1}{3}\left(\frac{1}{2}\dot \phi^2
+\frac{b^2}{\lambda^4}V\right)
\end{eqnarray}
where the derivatives are with respect to the  new time coordinate 
$\tilde \eta=a_0\lambda^2 \eta$ and the new scale factor is defined as 
$b(\tilde \eta)=a(\tilde\eta)/a_0$ with $a_0=a(0)$.
The solution of this equation shows that after the inflationary
phase, the scalar field starts oscillating around the minimum of the
potential with damped amplitude.
Substituting in (\ref{master}) 
and taking $\chi=e^{G/2}$ for this particular case, we obtain:
\begin{eqnarray}
\left(\frac{d^2}{d\tilde\eta^2}+\kappa^2+\frac{i}{\lambda^2}
\frac{d}{d\tilde\eta}(be^{G/2})+
\frac{b^2}{\lambda^4} e^{G}\right)f_{\kappa l}(\tilde\eta)=0
\label{master2}
\end{eqnarray}
with $\kappa=p/(a_0\lambda^2)$. From this expression we see that
when the scalar interaction is switched off, even in the expanding
background, there is no particle production. This is not surprising since
in that case, the equations of motion (in the masless limit) are conformally invariant.
 Following \cite{Baacke,Mostepanenko}
we can calculate the occupation number:
\begin{eqnarray}
N_{\kappa l}(\tilde T)=\frac{1}{4\kappa}\left(2 \kappa
+i[\dot f^*_{\kappa l}(\tilde T) 
f_{\kappa l}(\tilde T)\right.
- f^*_{\kappa l}(\tilde T)
\dot f_{\kappa l}(\tilde T)]
-\left.\frac{2}{\lambda^2}be^{G(\tilde T)/2}
\vert f_{\kappa l}(\tilde T)\vert ^2 ]\right)
\label{occupation}
\end{eqnarray}

In order for the particle number to be well defined, we will evaluate
it when the interaction is vanishingly small, that is, for large values 
of $\tilde T$. 
Here $f_{\kappa l}$ is a solution of equation (\ref{master2})
with initial conditions $f_{\kappa l}(0)=1$ and 
$\dot f_{\kappa l}(0)=-i\kappa$ which corresponds to a plane wave for
$\tilde \eta\leq 0$. In order to define the initial vacuum
 at $\tilde \eta=0$, we have taken the inflaton to be at the minimum of
the potential at that moment ($\phi(0)=1$), which implies that 
the interaction term vanishes, i.e, $e^{G(\phi=1)/2}=0$. 
 According to the
definition of $b$ we also have $b(0)=1$. We have chosen $\dot \phi(0)=1.8$
in our numerical computations which corresponds to  an initial amplitude
of the inflaton oscillations around $0.06M_p$.
\begin{figure}
\vspace{3cm}
\hspace{2cm} $N_{\kappa l}$
\vspace{-5cm} 
\begin{center}
\mbox{\epsfysize=9cm\epsfxsize=9cm
\epsffile{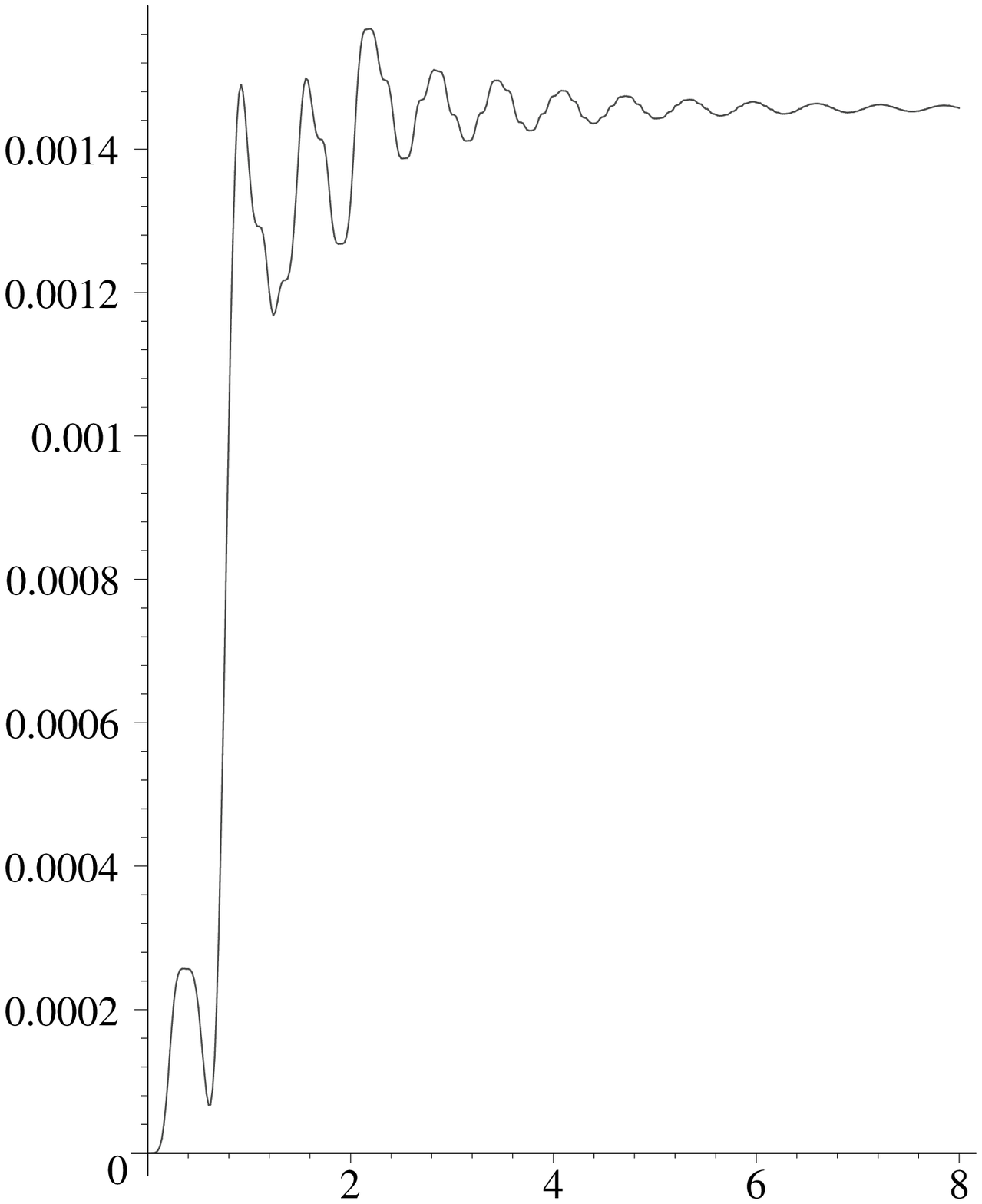}}
\end{center}
\vspace {-1.5cm}

\hspace{7cm} $\tilde \eta$

\leftskip 1cm
\rightskip 1cm

\vspace{.5cm}

{\footnotesize
{\bf Figure 1.-}Number of helicity $l=\pm 3/2$ gravitinos ($N_{\kappa l}$)
as a function of time $\tilde \eta$ for $\kappa=5$} 
\end{figure}

In Fig.1, the 
behaviour of $N_{\kappa l}$ has
been plotted as a function of time. We see how the production takes place
in a few inflaton oscillations and that asymptotically the
number of particles created is a well-defined quantity since the interaction vanishes.
The results for the spectra in the expanding background can be found in Fig.2. 
Notice that
we have not considered the backreaction effect of the produced particles
on the scalar field evolution. Because of the expansion of the universe, 
the production is significantly reduced with respect to the
flat space case. As expected \cite{Linde} the
resonance structure is affected by the expansion and the Pauli
limit is not saturated, but the number 
of particles that are produced is not negligible.

\begin{figure}
\vspace{-.5cm}
\begin{center}
\mbox{\epsfysize=9cm\epsfxsize=9cm
\epsffile{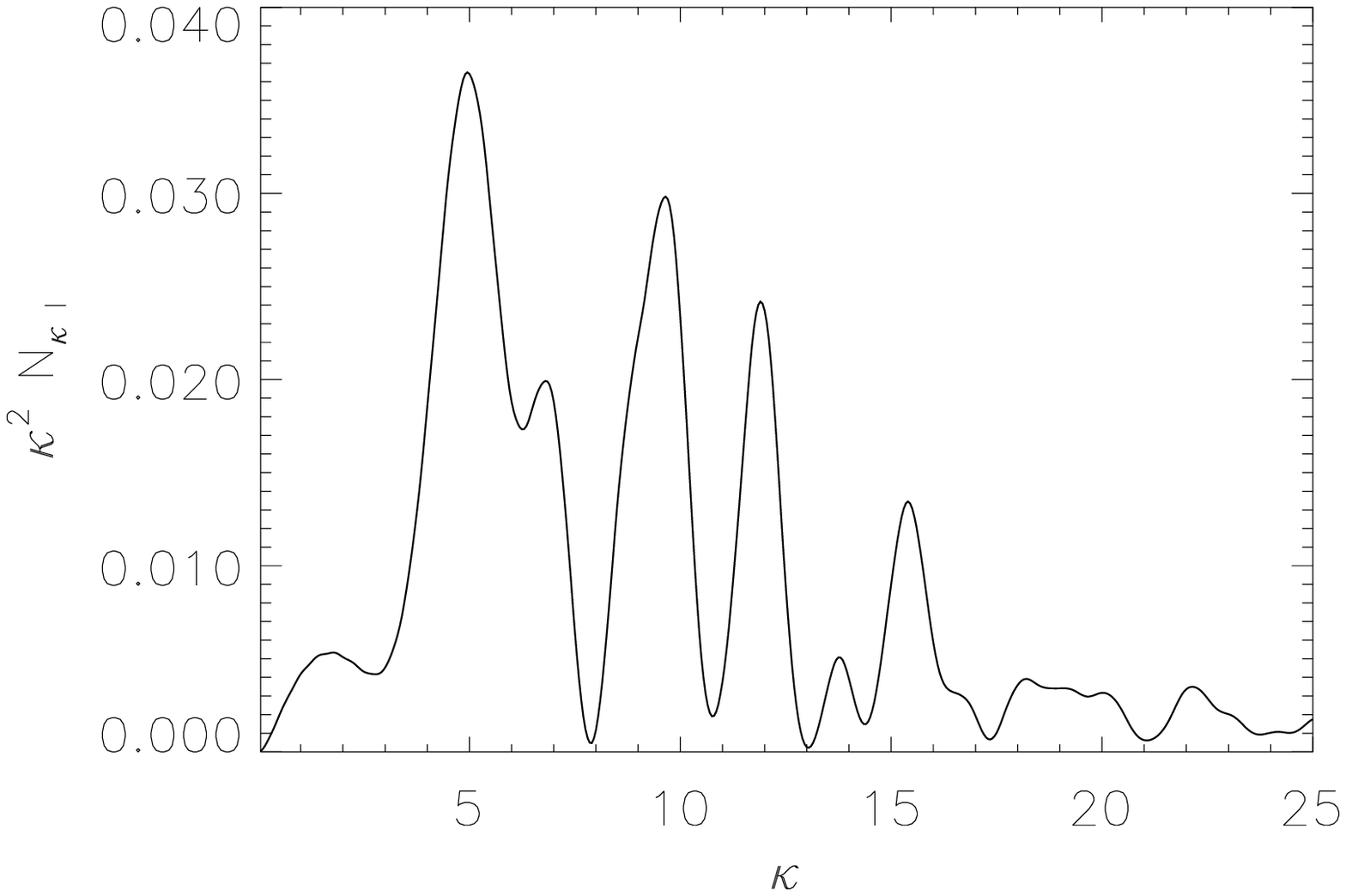}}
\end{center}
\vspace {-.7cm}
\leftskip 1cm
\rightskip 1cm
{\footnotesize
{\bf Figure 2.-}Number density of helicity $l=\pm 3/2$ gravitinos ($\kappa^2
N_{\kappa l}$)  against
$\kappa$. } 
\end{figure}
We expect modifications in the spectrum for different
initial conditions, although the total number of particles created 
is not very sensitive to these changes.
From Fig. 2, we can estimate a lower bound to  the total number density of 
gravitinos with both helicities as:
\begin{eqnarray}
n_{pre}(\eta)&=&\frac{1}{\pi^2 a^3(\eta)}\int_{p_{min}}^{\infty} N_{pl}p^2 dp
=
 \frac{a_0^3\lambda^6}{\pi^2 a^3}\int_{\kappa_{min}}^{\infty} 
N_{\kappa l}\kappa^2 d\kappa
\label{numer}
\end{eqnarray}
with $p_{min}=2\pi H(\eta)$. Since today, ($\eta_0, a(\eta_0)=1$)
$H<<M a_0\lambda^2$, we get: 
$n_{pre}(\eta_0)\geq a_0^3 10^{28}GeV^3$. To be compared with
the number density of a thermal distribution of helicity $\pm 1/2$
 gravitinos as
estimated in \cite{Pagels} (the helicity
$\pm 3/2$ could be even less dense): $n(\eta_0)\simeq 10^{-40}GeV^3$.
The comparison depends on the dilution term $a_0^3$, i.e. on
the scale factor at the end of inflation, and it shows that, for example, 
for a typical
value \cite{Kolb} $a_0\simeq 10^{-26}$,
the vacuum fluctuation production is suppressed with respect
to the thermal distribution by a factor $\geq 10^{-10}$. 
The corresponding cosmological consequences have been studied in  
\cite{LiEll}. Comparing with the entropy density today we get:
$n_{pre}/s \geq 10^{-12}$.

If we compare these results with the nucleosynthesis bounds
given in the introduction, we see that the preheating 
production is compatible with them for a  mass
of (unstable) gravitino 
$m_{3/2}> \Od(10 TeV)$ for this particular model.
This implies that gravitino masses $100 MeV< m_{3/2}<10 TeV$ would
be forbidden. On the other hand, if gravitinos are stable particles, then
in order not to overclose the universe we get: $m_{3/2}<1 TeV$. These bounds should be
taken with caution, as the results strongly depend on the model parameters. 
 
In
the standard scenario of reheating for this model, the reheating temperature
is very low $T_R\simeq 10^5 GeV$ which implies that the perturbative gravitino production, 
that can take place either from $2 \rightarrow 2$ proccesses involving  gauge bosons
and gauginos or from direct inflaton decay \cite{Sarkar}, 
is very small $n_{reh}/s\simeq \Od(10^{-17})$. Comparing with 
the preheating results above, we see that the difference between the two 
mechanisms is apparent in these estimations.  

{\bf Acknowledgements:} I thank A. Mazumdar for useful discussions. This 
work has been supported by SEUID-Royal Society and
CICYT-AEN97-1693 (Spain).

\end{document}